# New cosmic ray observations at Syowa Station in the Antarctic for space weather study

C. Kato[1], W. Kihara[1], Y. Ko[1], A. Kadokura[2,3,6], R. Kataoka[2,6], P. Evenson[4], S. Uchida[1], S. Kaimi[1], Y. Nakamura[5], H.A. Uchida[2,6], K. Murase[6], and K. Munakata[1]

[1] Department of Physics, Faculty of Science, Shinshu University, Matsumoto, Nagano, JAPAN
   e-mail: ckato@shinshu-u.ac.jp
[2] National Institute of Polar Research (NIPR), Tachikawa, Tokyo, JAPAN
[3] Polar Environment Data Science Center, Joint Support-Center for Data Science Research, Research Organization of Information and Systems, Tachikawa, Tokyo, JAPAN
[4] Bartol Research Institute, Department of Physics and Astronomy, University of Delaware, Newark, DE, Institute USA
[5] Key Laboratory of Particle Astrophysics, Institute of High Energy Physics, Chinese Academy of Sciences, Beijing 100049, China
[6] Department of Polar Science, The Graduate University for Advanced Studies, SOKENDAI, JAPAN

**ABSTRACT**

Muon detectors and neutron monitors were recently installed at Syowa Station, in the Antarctic, to observe different types of secondary particles resulting from cosmic ray interactions simultaneously from the same location. Continuing observations will give new insight into the response of muon detectors to atmospheric and geomagnetic effects. Operation began in February, 2018 and the system has been stable with a duty-cycle exceeding 94%. Muon data shows a clear seasonal variation, which is expected from the atmospheric temperature effect. We verified successful operation by showing that the muon and neutron data are consistent with those from other locations by comparing intensity variations during a space weather event. We have established a web page to make real time data available with interactive graphics (http://polaris.nipr.ac.jp/~cosmicrays/).

**Key words.** Cosmic Ray–Muon Detector–Neutron Monitor

## 1. Introduction

Solar cycle 24 was the weakest of those in the space age and the current solar minimum is expected to show very weak solar activity, similar to the deep solar minimum in 2009 (Smith et al. (2014), https://solarscience.msfc.nasa.gov/SunspotCycle.shtml). Since the solar modulation of galactic cosmic rays (GCR) should be very weak during the minimum around 2020, there is a good chance to measure another record-high GCR intensity (Miyake et al. (2017)). Continuous observations of high energy cosmic ray intensity over additional solar cycles contributes to research on space weather and space climate, and especially radiation dose for spacecraft design and human activity in high altitudes and in space.

Applied science pertaining to solar energetic particles (SEP) during the comimg solar maximum will also be important since the highly energetic component of SEP, responsible to the ground level enhancement (GLE), causes a significant radiation dose at flight altitudes. Real-time data acquisition is important as data from ground-based neutron monitors are utilized to run a warning system of radiation dose for aircrews (Kuwabara et al. (2006)), in the International GLE Database (Usoskin et al. (2020), https://gle.oulu.fi), and WASAVIES (Kataoka et al. (2014), Kataoka et al. (2018), Sato et al. (2018))).

Major ground based Cosmic Ray (CR) observations for space weather study have been performed with Neutron Monitors (NM) and Muon Detectors (MD), especially network observations such as the Space Ship Earth (SSE) project using 11 NM, the entire world-wide NM network [1,2], and the Global Muon Detector Network (GMDN) project using 4 multi-directional MD (e.g. Bieber et al. (2004) , Kuwabara et al. (2009)). Muon and neutron networks observe GCR of different median rigidities and therefore reveal rigidity dependencies of GCR fluxes in space weather and space climate events. The median primary rigidity ($P_m$) of a detector is defined as an rigidity above/below which the detector has 50 % of its total response to primary GCR. Average $P_m$ calculated with Dorman's response function for the SSE network is about 10 GV, while the representative $P_m$ calculated with Murakami's response function for the GMDN is about 50 GV. By analyzing NM and MD data together, we can study the GCR modulation in a wide range of $P_m$ (Moraal et al. (2000), Clem and Dorman (2000), Murakami et al. (1979)).

Different responses of NM and MD to atmospheric and geomagnetic effects need to be carefully examined to studying rigidity dependence of GCR modulation in space. The best way to understand the different responses is simultaneous observations with NM and MD beneath the same atmosphere at the same location in the Arctic and/or Antarctic. The main reason to chose polar region is the reduced deflection of GCRs in the geomagnetic field. NM and MD installed at a location in the Arctic/Antarctic can observe GCRs arriving from similar direction outside the magnetosphere because GCR, as charged particles, tend to arrive along geomagnetic field lines. On the contrary, the trajectories of GCR low and high rigidity observed at low latitude are more significantly and differently deflected by the geomagnetic field.

Historically, Japanese Antarctic Research Expedition (JARE) projects performed CR observation with NMs in the Antarctic until early 1960s (e.g. Kitamura and Kodama (1961)). CR observations resumed in February 1st, 2018 at Syowa Station in the Antarctic after nearly six decades. These new observations are expected to play an important role in analyzing the combined data from GMDN and world wide NM network including the SSE. It is the purpose of this paper to present the first light data from observed atmospheric and space weather events during the solar minimum. After describing general aspects of measurements and calibrations in Section 2, we describe the new observation system at Syowa Station in Section 3 and present first light data in Section 4. Concluding remarks will be summarized in Section 5.

---

[1] http://cidas.isee.nagoya-u.ac.jp/WDCCR
[2] http://nmdb.eu

## 2. Ground level cosmic ray observations

GCRs are high energy charged nuclei, mainly protons, arriving at Earth from outer space. GCRs entering the upper atmosphere generate secondary particles through the interaction with atmospheric nuclei. All ground based CR detectors detect secondary CRs; NM and MD detect secondary neutrons and secondary muons, respectively. Design and construction of the detectors at Syowa station is described in some detail in Section 3, but first we present a brief, functional overview.

A specific NM design, termed NM64 is constructed around a Proportional Counter Tube (PCT) filled with $^{10}BF_3$ gas. Boron nuclei capture neutrons by a nuclear fission reaction [$^{10}$B(n,$\alpha$)$^7$Li] and the energy released is detected by amplifying the ionization in the gas (e.g. Simpson (2000)). The PCT is enclosed in three layers of material (see figure 4) called, from inside out, moderator, absorber/producer, and reflector. The moderator is a thin polyethylene layer which decelerates neutrons so that they can be captured by boron nuclei. The absorber/producer consists of lead rings that emit low energy (few MeV) neutrons when struck by higher energy neutrons in secondary CR. The reflector, made of polyethylene, prevents thermal neutrons from escaping before they can be moderated and absorbed. The reflector also excludes environmental low energy neutrons and other background radiation such as gamma rays, electrons and positrons.

MD used for the GMDN consist of PCT filled with argon and methane or plastic Scintillation Detectors (SciD), both of which are commonly used for detecting various types of charged particles. As CR detectors, PCTs and SciDs are arranged in orthogonal layers, vertically separated and used in coincidence to determine the incident direction of muons. [3, 4].

Several things must be taken into account when analyzing data from NM and MD, most importantly response functions, atmospheric effects, and geomagnetic effects. Details of these can be found in books such as Dorman (2004) and Dorman (2009). The response function gives the counting rate of a detector at a specific location on the globe for some specific direction as a function of the rigidity spectrum of primary GCR, and is quite different for NM and MD. The functions derived by Nagashima et al. (1989) and Murakami et al. (1979) are used in this paper for analyzing NM and MD data, respectively. The function presented in Clem and Dorman (2000) is also widely used for analyzing NM data. Recently, the response function based on the Monte Carlo simulations of hadronic shower development in the atmosphere has been developed (Mishev et al. (2020), Mangeard et al. (2016)).

While the response function gives the average CR flux to be observed by a particular detector under the average pressure ($P$) and temperature ($T$) of the atmosphere above the detector, the observed count rate ($n$) is subject to temporal variation due to variation of the pressure $P$ and temperature $T$. This atmospheric effect can be expressed, as

$$\log(\Delta n/n) = \beta \Delta P + \int_0^{h_{obs}} \alpha(h) \Delta T(h) dh \qquad (1)$$

---

[3] http://cosray.shinshu-u.ac.jp/crest/DB/Documents/Docs/DetectorDescription_NGY.pdf
[4] http://cosray.shinshu-u.ac.jp/crest/DB/Documents/Docs/DetectorDescription_KWT.pdf

where $\Delta n$ and $\Delta P$ are deviations of $n$ and $P$ from averages, respectively, $\Delta T(h)$ is the deviation of the temperature at an atmospheric depth $h$ in g/cm$^2$ and $\beta$ and $\alpha$ are coefficients of the pressure effect and the partial temperature effect, respectively.

The physical cause of the pressure effect is the energy loss and absorption of muons and neutrons in the atmosphere, both of which increase with increasing atmospheric pressure. It results in a negative $\beta$ in the first term on the right hand side of Eq. (1) for both of neutron and muon observations. The atmospheric pressure effect simply depends on the matter above the detector system. Strong winds, observed at Syowa Station occasionally, cause a barometer to detect a lower pressure due to the Bernoulli effect. It is necessary to perform close analysis of this and other local effects.

The physical cause of the temperature effect in the second term is the atmospheric expansion caused by the temperature increase, which works in two ways, 1) more CR muons would be produced at higher altitude, 2) more pions and kaons would decay into high energy muons rather than interact with atmospheric nuclei due to the longer interaction mean free path. The first effect causes a decrease of low energy muon count rate because more low-energy muons decay before reaching detector due to the increased path to the detector. This effect results in a negative $\alpha$ in Eq. (1). However, this is not the case for muons with higher energy, where the increased production rate dominates and results in a positive $\alpha$. The temperature effect on the sea level muon count rate appears predominantly as a negative effect in surface detector, while it appears as a positive effect in the observation of higher energy muons with underground detectors. On the other hand the NM count rate, is almost free from this effect.

The major geomagnetic effects on primary GCR are the asymptotic direction and the geomagnetic cut-off rigidity ($P_c$). The asymptotic direction from which GCRs arrive outside the magnetosphere strongly depends on the rigidity of GCRs, the central direction of Field Of View (FOV) and the geographic location of each detector. The asymptotic viewing direction of each detector can be calculated by tracing back orbits of antiparticles ejected toward the central direction of FOV of the detector into a model magnetosphere. By tracing back of orbits with different rigidity we also determine $P_c$ as the rigidity below which particles are trapped inside the magnetosphere.

## 3. CR detectors at the Syowa Station

Syowa Station, established in 1957 on East Ongul Island, Lützow Holm Bay, has since been used as a year-round scientific research home base. New CR detectors were installed by the 59th Japanese Antarctic Research Expedition (JARE59) in late 2017. The detectors, assembled in two refrigeration containers, started operation in February 1st, 2018. The containers are located at 69.01°S, 39.59°E and 24.7m above sea level. Each container is mounted on four concrete supports, 1 m above the ground surface, as shown in figure 1(a), to avoid heavy snow buildup around the container. The picture was taken in winter season of 2018 showing that the snow buildup is minor – partly due to the strong wind.

As shown in figure 1(b), a 3NM64 was installed in each of the two containers No.1 and No.2, while a MD was constructed only in container No.2, beneath the 3NM64. Configuration of the detectors in the container No.2 is shown in figure 2.

The NM is an omni-directional detector, which observes CRs arriving from the vertical direction on average. Three NM64 tubes are set on top of the steel frame. The absorber/producer for each tube is formed by 18 lead rings, each with a mass of 90 kg (see figure 3). A schematic view of a 3NM64 is shown in figure 4. The lead also shields the MD from the soft component radiation such as electrons, positrons and gamma rays.

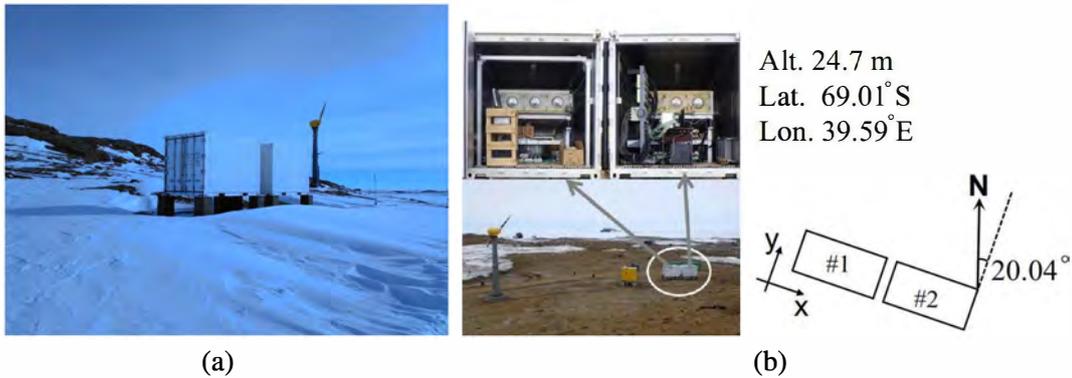

**Fig. 1.** Containers at Syowa Station. (a) Photo showing the environment of the containers in winter 2018. (b) Photo showing CR detectors installed inside containers. Location and orientation of the containers are also indicated.

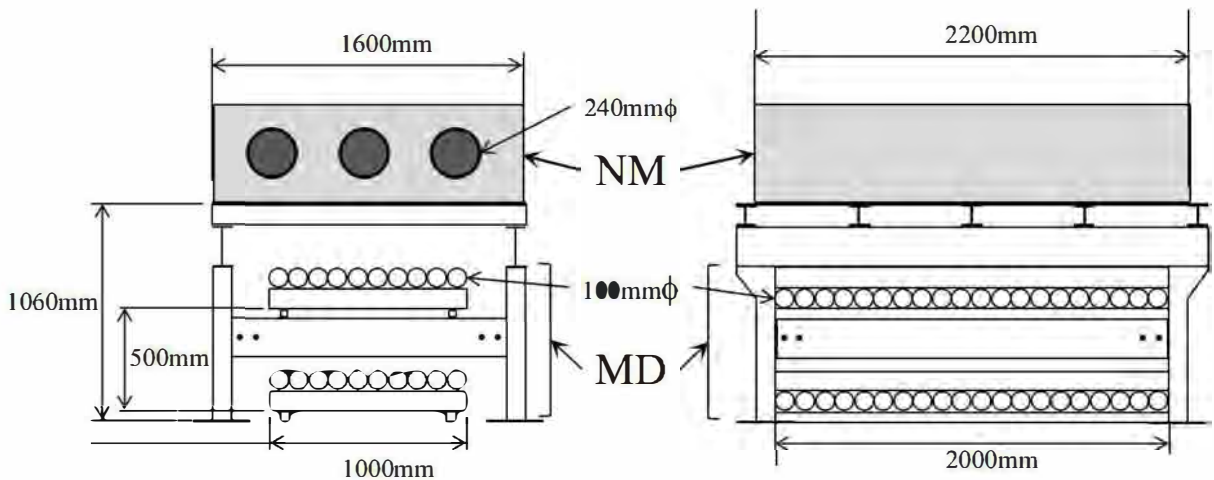

**Fig. 2.** Schematic view of detectors assembled on a steel frame in the container No.2.

Syowa MD comprises four horizontal layers of 10 cm diameter cylindrical PCTs. In the top and third layers, the 2 m long tubes are aligned along the X ("East-West") axis, while in the other two layers, 1 m long tubes are orthogonally aligned along the Y ("North-South") axis. The Y-axis of the MD is actually oriented 20.04 degrees eastward of true north (see figure 1(b)). Currently, only one MD system is installed at Syowa Station. This arrangement and 4-fold coincidence technique allow us to determine the incident direction of each CR muon (figure 5(a)). The muon count rate is recorded in each of 23×19=437 directional channels. To analyze Syowa MD data with better statistics, 437 directional channels are combined into 9 channels, which are equivalent to those available from 3 × 3 arrays of 50 cm × 50 cm square detectors on the upper and lower layers (see figure 5(b)).

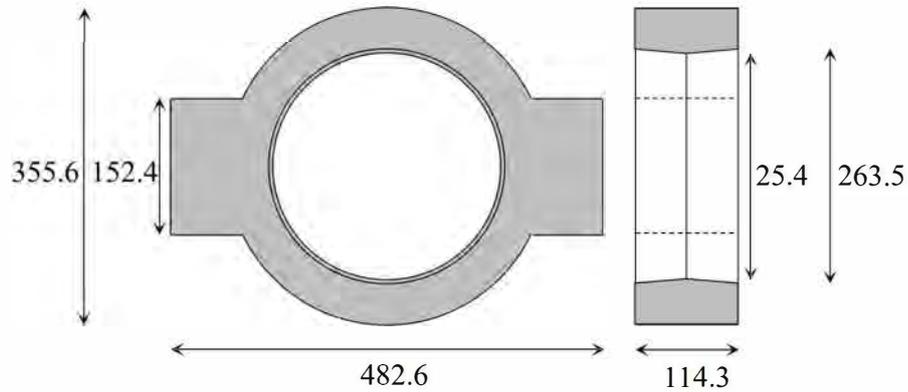

**Fig. 3.** Lead ring for NM. Mass is 90 kg and dimensions are in mm.

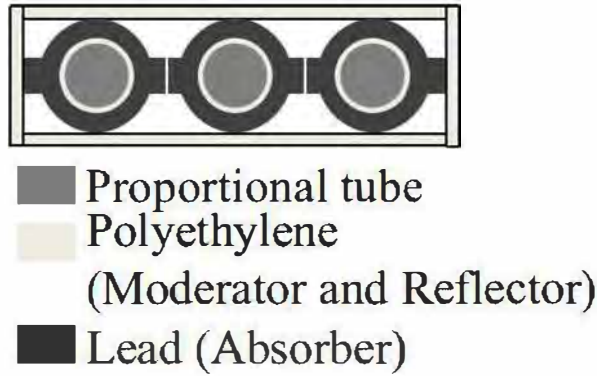

**Fig. 4.** Schematic view of a 3NM64.

The characteristics of Syowa NM and MD are summarized in Table 1. Median rigidity $P_m$ (and asymptotic directions at $P_m$) of primary GCR is calculated by using response functions by Nagashima et al. (1989) and by Murakami et al. (1979) for NM and MD, respectively. The response function for NM varies depending on the solar activity, because the energy spectrum of primary GCRs becomes softer (harder) in the solar activity minimum (maximum) period. Due to this, $P_m$ also slightly varies depending on the solar activity. $P_m$ for Syowa NM in Table 1 is calculated for the solar minimum period. Asvestari et al. (2017) has recently proposed to use the "effective energy" instead of $P_m$. The response function and $P_m$ for MD, on the other hand, are much less dependent on the solar activity.

The GMDN consists of four multi-directional MDs located at Nagoya (Japan), Hobart (Australia), São Martinho (Brazil) and Kuwait City (Kuwait). Asymptotic viewing directions of the GMDN and Syowa MD are shown in figure 6. Each small symbol indicates the asymptotic direction of primary GCRs with the median primary rigidity ($P_m$) observed in each directional channel of GMDN and Syowa MD, while large solid symbols indicate the direction of vertical channel of each MD. The viewing direction of the Syowa NM is also plotted in the figure 6 as a blue cross. The track through each solid symbol represents the spread of viewing directions corresponding to the central 80 % of the rigidity response. This map demonstrates a possibility to make the normalization among existing GMDN detectors better by using the viewing directions of Syowa MD overlapping with those of other MDs. For developing such normalization and also for the integrated analyses of NM and GMDN data, simultaneous observations by MD and NM at the Syowa CR observatory is expected to play a key role.

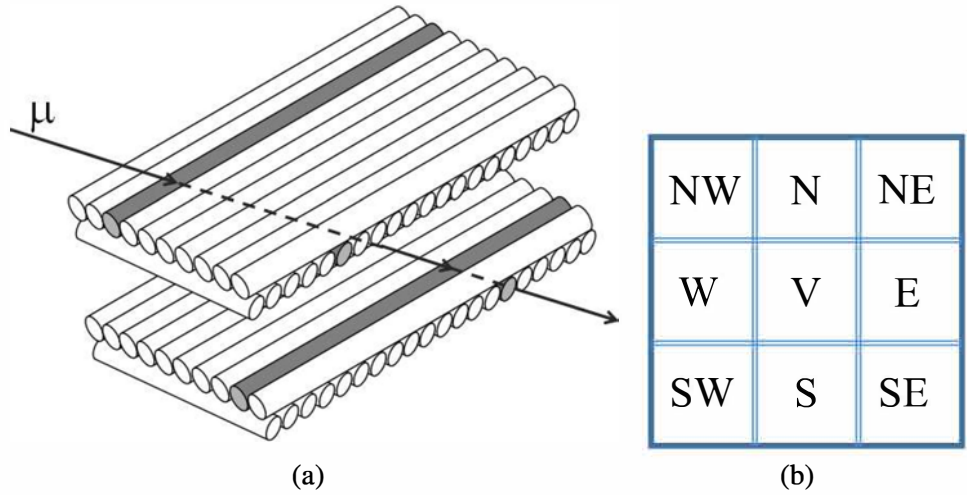

(a)                                                    (b)

**Fig. 5.** (a) Schematic view of 4-fold coincidence of MD. (b) Combined 9 directional channels for better statistics. These channels are equivalent to those available from 3x3 arrays of 50cmx50cm square detectors on the upper and lower layers.

**Table 1.** Characteristics of Syowa NM and MD. $P_c$ of NM is calculated for vertical direction.

| Directional channel | Cutoff Rigidity $P_c$[GV] | Median Rigidity $P_m$[GV] | Asymptotic directions@$P_m$ | Correction coefficient $\beta$ [%/hPa] | Average Count rate $\times 10^4$[counts/hr] |
|---|---|---|---|---|---|
| Syowa NM |  |  |  |  |  |
| omni | 0.4 | 16.3 | 30°S38°E | -0.76 | 31.8 (Total) |
|  |  |  |  |  |  |
| Syowa MD |  |  |  |  |  |
| V |  | 53.6 | 56°S53°E | -0.16 | 26.9 |
| N |  | 58.5 | 29°S35°E | -0.16 | 3.8 |
| S |  | 58.5 | 70°S115°E | -0.16 | 3.8 |
| E |  | 58.5 | 32°S81°E | -0.16 | 6.4 |
| W |  | 58.5 | 60°S8°W | -0.16 | 6.4 |
| NE |  | 63.6 | 16°S60°E | -0.15 | 1.0 |
| NW |  | 63.6 | 34°S4°E | -0.16 | 1.0 |
| SE |  | 63.6 | 42°S110°E | -0.16 | 1.0 |
| SW |  | 63.6 | 80°S77°W | -0.16 | 1.0 |

Data are recorded in a PC at the site and transferred to a data server at Shinshu University in Japan via internet. Transferred data are automatically corrected for the atmospheric pressure effect and then stored in a database. Data are accessible via a web page[5]. This web page is designed to provide NM and MD data graphically. Count rates in the NM and the vertical directional channel of MD can be plotted for a desired time period. Several functions are implemented to the web page, such as 1) selectable time resolution, 2) zooming vertical axis, 3) reading values on the plot, and 4) downloading data as a text file. It is planned to enhance this web page for researchers to analyze space weather events. The data will soon become available in the existing international databases (see footnotes in Section 1).

---

[5] http://polaris.nipr.ac.jp/~cosmicrays/

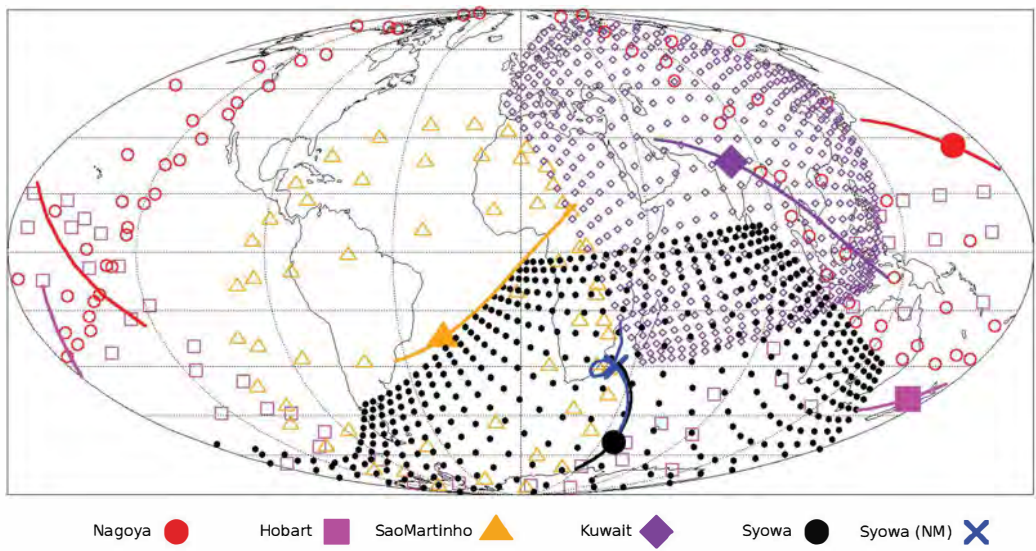

**Fig. 6.** Asymptotic viewing directions of the GMDN and MD+NM at Syowa Station. Each small symbol shows the asymptotic direction of primary GCR having the median primary rigidity ($P_m$) observed in each directional channel of GMDN and Syowa MD. Each large solid symbol indicates the viewing direction of the vertical channel in each MD, while a large cross symbol indicates a viewing direction of Syowa NM. The track through each solid symbol represents the spread of viewing directions corresponding to the central 80 % of the rigidity response.

## 4. Preliminary observations and data analysis

The detectors at Syowa Station have operated since February, 2018, with a duty cycle through October 31st, 2019 of 94% for NM and 99% for MD. In this report, we present our preliminary analysis of the data recorded between February 2018 and October 2019. The top two panels in figure 7 display the percent deviation from the average of the pressure corrected daily mean count rate (in the vertical channels) of Nagoya MD (NGY_MD_V) and Syowa MD (SYO_MD_V). The lower three show Syowa NM (SYO_NM), Polar Bare NM (SOPB) and South Pole NM (SOPO) in the same format. SOPO and SOPB are NMs with/without lead absorber/producer installed at the South Pole. The typical statistical error of each data point in this figure is less than 0.1%. Geographic locations and the geomagnetic cut-off rigidity ($P_c$) calculated for each detector are listed in Table

2. Significant seasonal variations due to the atmospheric temperature effect are clearly seen in MD data. The variations are out of phase in NGY_MD_V and SYO_MD_V because of their locations in opposite hemispheres. NM data, on the other hand, show no such large seasonal variations as expected. The NM data from Antarctica show similar variations. The squared differences between SOPO and SYO_NM, between SOPB and SYO_NM and between SOPO and SOPB are shown in figure 8. Differences between SOPO and SYO_NM, and SOPB and SYO_NM are large in the beginning of the observation. Although there are still several periods showing large differences after 2019, the differences seem to be smaller. In 2018, the correlation coefficients between SYO_NM and SOPO is 0.31, and that between SYO_NM and SOPB is 0.67. In 2019, these correlation coefficients increase to 0.43, and 0.78, respectively. Since a heating system was installed in each container at Syowa Station to maintain a steady temperature in January, 2019, this could be a reason for the reduced difference in 2019. Another possible origin of the difference is the occasional influence on the measured atmospheric pressure by the Bernoulli effect from strong wind. We need to study such possible local effects in more detail.

**Table 2.** Geographic location and cut off rigidity of Nagoya MD, Syowa MD & NM, SOPO, and SOPB. $P_c$ of SyowaNM and MD is calculated for vertical direction.

| Station | Cutoff Rigidity $P_c$[GV] | Location | Altitude |
|---|---|---|---|
| Nagoya MD_V | 11.5 | 35.15°N136.97°E | 77m |
| Syowa NM,MD_V | 0.4 | 69. 01°S 39.59°E | 24.7m |
| SOPO&SOPB | 0.1 | 90°S | 2820m |

Mendoça et al. (2016) developed a correction method for the atmospheric temperature effect by calculating the second term on the right hand side of Equation (1), as

$$\int_0^{h_{obs}} \alpha(h)\Delta T(h)dh \Rightarrow \alpha_{MSS} \times \Delta T_{MSS} \tag{2}$$

where $\alpha_{MSS}$ is a single correction coefficient and $\Delta T_{MSS}$ is the deviation of the mass-weighted temperature from the average defined as

$$\Delta T_{MSS} = \sum_{i=0}^{n} \Delta T[h_i] \times \frac{x[h_i] - x[h_{i+1}]}{x[h_0]} \tag{3}$$

with $x[h_i]$ and $\Delta T[h_i]$ denoting the atmospheric depth and the temperature-deviations from the average at an altitude $h_i$ ($h_i = h_0$ at the ground level). The correction coefficient $\alpha_{MSS}$ is calculated from the correlation between the pressure corrected CR count rate and $\Delta T_{MSS}$. We apply this correction method to Syowa MD data by using $\Delta T_{MSS}$ derived from the Global Data Assimilation System (GDAS) by NOAA[6]. The temperature at 925 hPa is used for $T[h_0]$ instead of the ground level temperature. Figure 9 shows a scatter plot between $\Delta T_{MSS}$ and the pressure corrected daily mean count rate in SYO_MD_V. The correction coefficient $\alpha_{MSS}$ is calculated to be -0.289±0.004 [%/K] with the correlation coefficient of 0.94.

---

[6] https://www.ncdc.noaa.gov/data-access/model-data/model-datasets/global-data-assimilation-system-gdas

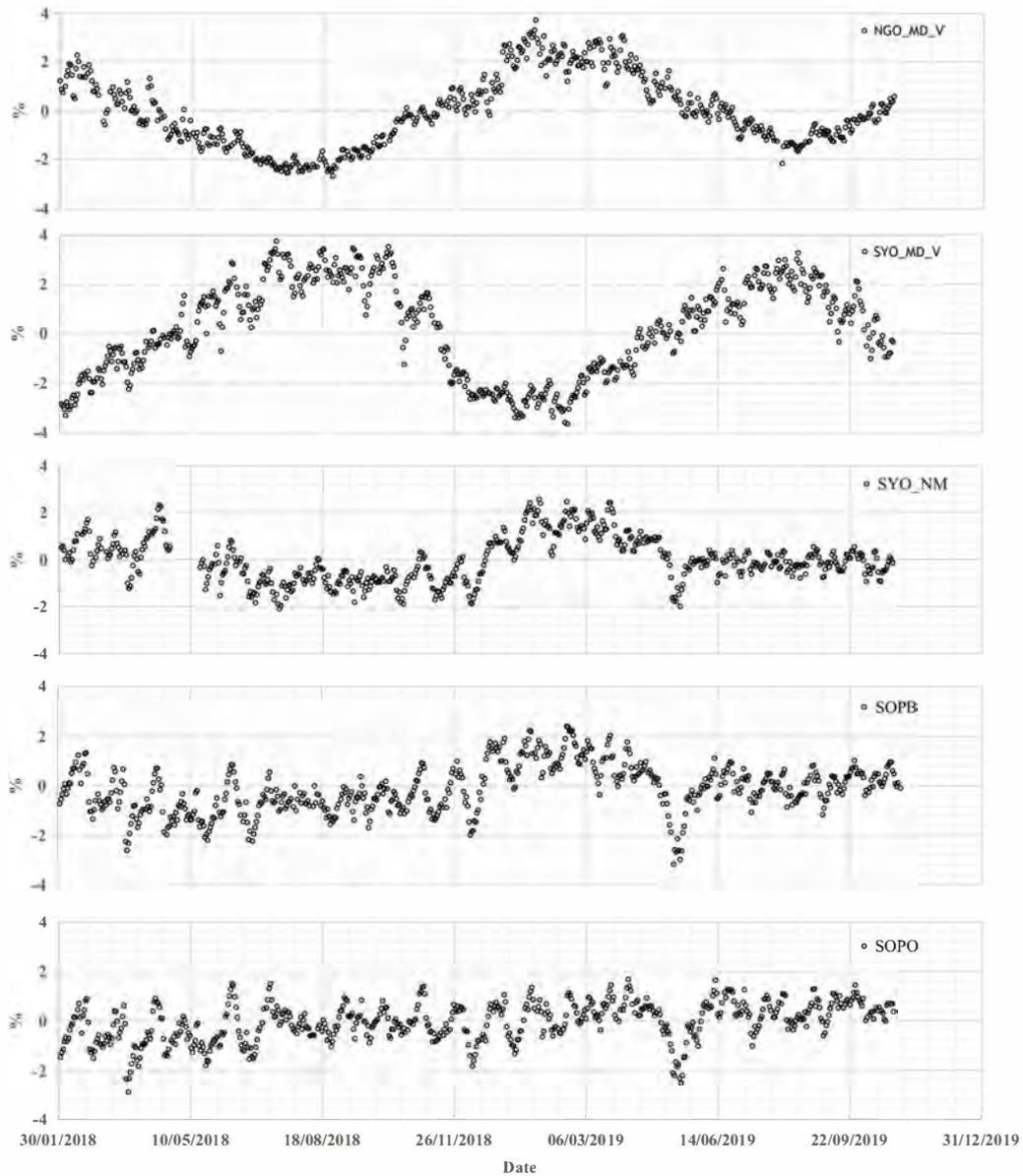

**Fig. 7.** Temporal variations of the pressure corrected daily mean count rates recorded by NMs and MDs. From top to bottom, each panel displays the percent deviation from the average of daily mean count rates in Nagoya MD vertical channel (NGY_MD_V), Syowa MD vertical channel (SYO_MD_V), Syowa NM (SYO_NM), Polar Bare NM (SOPB), and South Pole NM (SOPO), during a period between February, 2018 and October, 2019. Representative error of each data point is less than 0.1%. MD data in top two panels show clear seasonal variations due to the atmospheric temperature effect. Minor tick marks on the horizontal axis are indicated every 27 days.

Top panel of figure 10 displays the seasonal variation reproduced from $\alpha_{MSS}$ and $\Delta T_{MSS}$ (blue circles) together with the observed variation (black open triangles), while the bottom panel shows the observed variation corrected for the temperature effect. The seasonal variation of muon count rate is quantitatively evaluated by using a Monte-Carlo (MC) simulation code, PHITS version 3.20 (Sato et al. (2018)), for the particle transport in the atmosphere. We set a simple cylinder-shaped atmosphere with a radius of 100 km and a hight of 200 km, in which the altitude profile of the atmospheric density is set according to MSISE-00 model (Picone et al. (2002)). GCR protons are isotropically injected at the center of the top cylinder surface, and a power-law spectrum with the spectral index of -2.7 is assumed for the differential energy spectrum of GCRs. The simulated variation of the monthly mean count rate in SYO_MD_V is plotted in the top panel of figure 10 by red open triangles. The simulated variation is overall consistent with the observation.

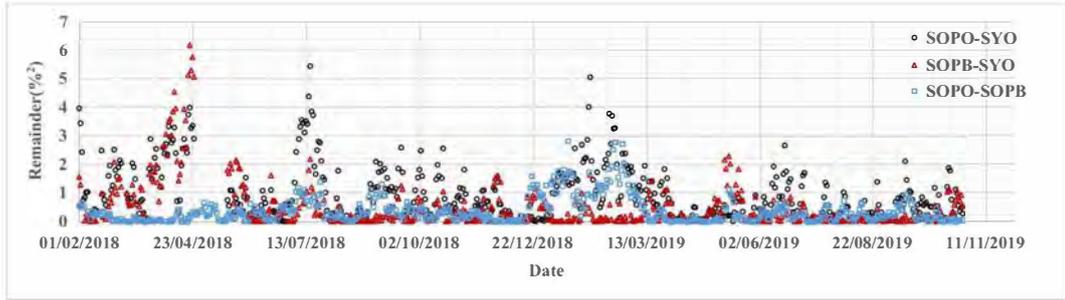

**Fig. 8.** The squared difference between variations observed by three NMs. (SOPO − SYO_NM)$^2$, (SOPB − SYO_NM)$^2$ and (SOPO − SOPB)$^2$ are shown by black open circles, red open triangles and blue open squares, respectively. Differences became smaller after 2019, except in several short periods. Minor tick marks on the horizontal axis are indicated every 27 days.

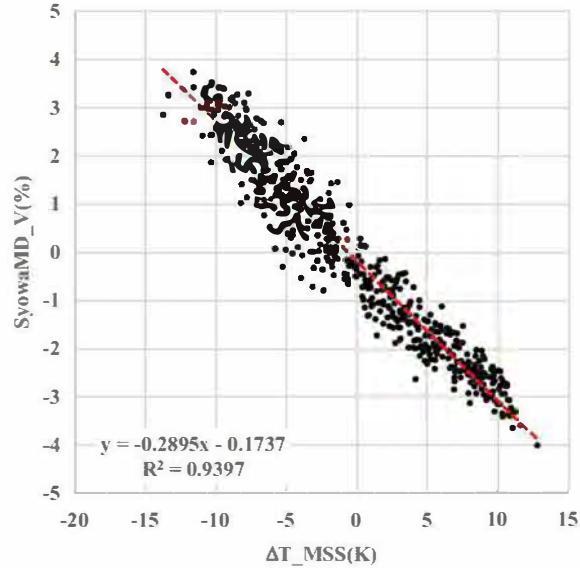

**Fig. 9.** A scatter plot between $\Delta T_{MSS}$ and the pressure corrected daily mean count rate in Syowa MD vertical channel (SYO_MD_V). The derived regression coefficient ($\alpha_{MSS}$) is -0.289±0.004 [%/K] and the correlation coefficient is 0.94.

Atmospheric temperatures at various altitudes are also measured twice a day at Syowa Station by Japan Meteorological Agency (JMA). We verified that the correction using the JMA temperature data is consistent with the correction using GDAS data mentioned above. Thus, GDAS and JMA data can be used complementarily.

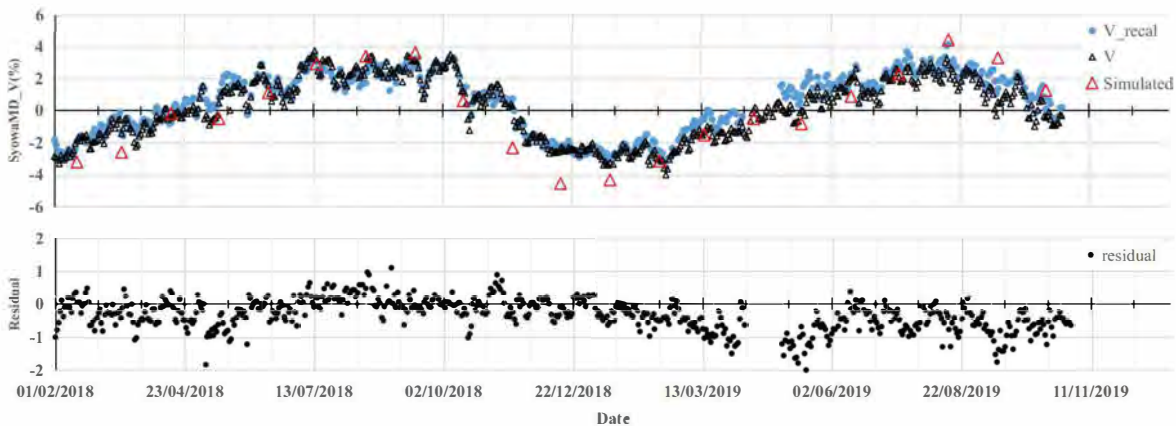

**Fig. 10.** Top: The pressure corrected daily mean count rate in Syowa MD vertical channel (SYO_MD_V) (open black triangle) and the variation reproduced from $\alpha_{MSS}$ and $\Delta T_{MSS}$ (blue circle) (see text). Open red triangles indicate the CR variation reproduced by the MC simulation of particle transport in the atmosphere (see text). Bottom: The observed variation corrected for the atmospheric temperature effect. Minor tick marks on the horizontal axis are indicated every 27 days.

The same correction method is also applied to Nagoya MD vertical channel (NGY_MD_V) data for comparison and the regression coefficient of -0.234 ± 0.002 [%/K] is obtained as shown in figure 11. The observed and reproduced seasonal variations in NGY_MD_V are shown in the top panel of figure 12 and the variation corrected for temperature effect is shown in the bottom panel. Although these results need to be examined in more detail, the correction method again appears like working reasonably well.

Two particularly interesting events were observed by the Syowa MD and Syowa NM. One is related to the Stratospheric Sudden Warming (SSW) in September, 2019. Figure 13 shows the daily mean count rate in SYO_MD_V together with $\Delta T_{MSS}$ calculated from Eq. (3), both from the temperature below (left panel) and above (right panel) the 200 hPa altitude equi-pressure surface (henceforth called the 200 hPa altitude). The SSW period of about 20 days is indicated by red rectangles. During the SSW, the SYO MD V rate is clearly suppressed, although the details are difficult to quantify due to the overall seasonal trend. A corresponding feature appears in the calculated temperature at the higher altitude, as plotted on the inverse scale. At lower altitude there is also a deviation in the calculated temperature, but that is in a direction opposite to the general trend of the data. After the SSW, SYO_MD_V varies in correlation with $\Delta T_{MSS}$ both above and below the 200 hPa altitude, showing a typical seasonal variation. Since the atmospheric temperature effect on muon rate is an integrated effect through the atmosphere at various altitude, SYO_MD_V data in figure 13 indicates that the SSW is caused by the warming over a wide range of high altitude atmosphere above the 200 hPa altitude and not due to the local warming/cooling of the atmosphere.

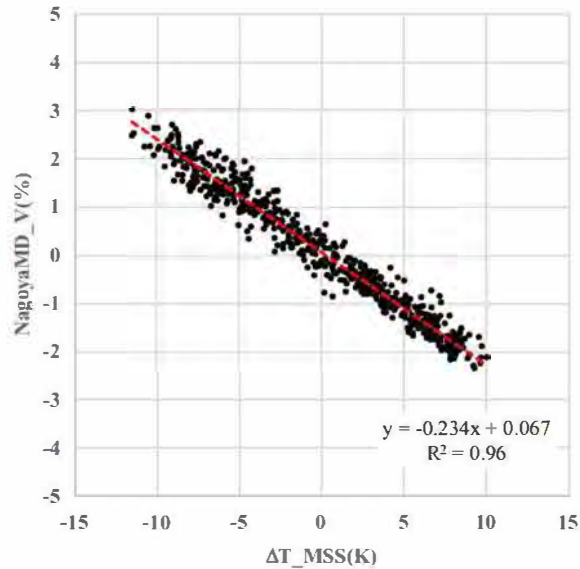

**Fig. 11.** A scatter plot between $\Delta T_{MSS}$ and the pressure corrected daily mean count rate in Nagoya MD vertical channel (NGY_MD_V). Derived regression coefficient ($\alpha_{MSS}$) is -0.234±0.002 [%/K] and the correlation coefficient is 0.96.

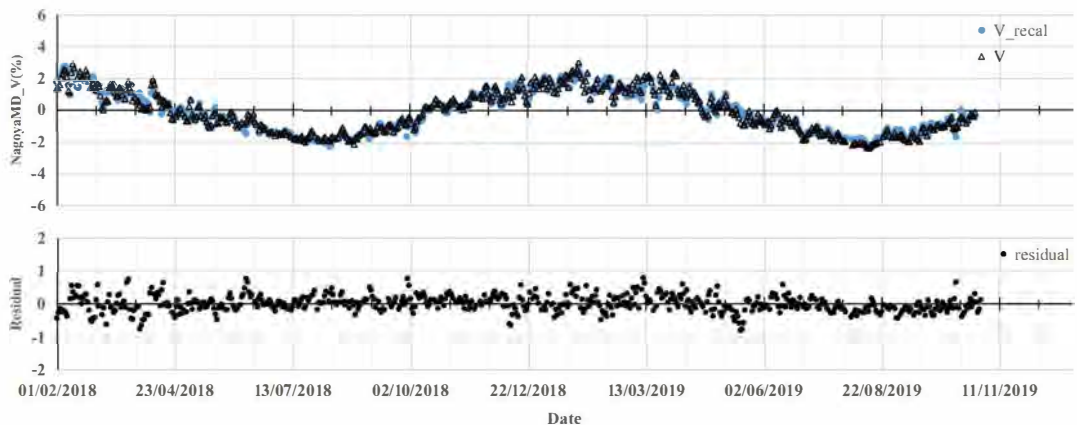

**Fig. 12.** Top: Daily average count rate variation of Nagoya MD (open triangle) and reproduced variation by atmospheric temperature (blue circle). Bottom: The observed variation corrected for the atmospheric temperature effect. Minor tick marks on the horizontal axis are indicated every 27 days.

Another space weather event was recorded in August, 2018 (Chen et al., 2019). Following a weak interplanetary shock on 25 August, 2018, a Magnetic Flux Rope (MFR) caused an unexpectedly large geomagnetic storm. It is likely that this event became geoeffective because the MFR was accompanied by a corotating interaction region and compressed by high-speed solar wind following the MFR. Abunin et al. (2020) analyzed a Forbush decrease observed during this event by the world-wide NM network and reported that this event is accompanied by the enhanced GCR anisotropy (see also Gil et al. (2018)). By analyzing the GMDN data recorded during this event, we found that a significant increase of the GCR density (omnidiectional intensity) exceeding the unmodulated level before the shock was observed for six hours between 04:00 and 09:00 on 26 August, near the trailing edge of the MFR. Significant north directed anisotropy was also found during the same six hours(Kihara et al. (2021)). Although it is out of scope of this paper to describe our analysis of this event, observed by the GMDN in detail, we show how the data recorded by Syowa NM and MD are consistent with the observation by the GMDN.

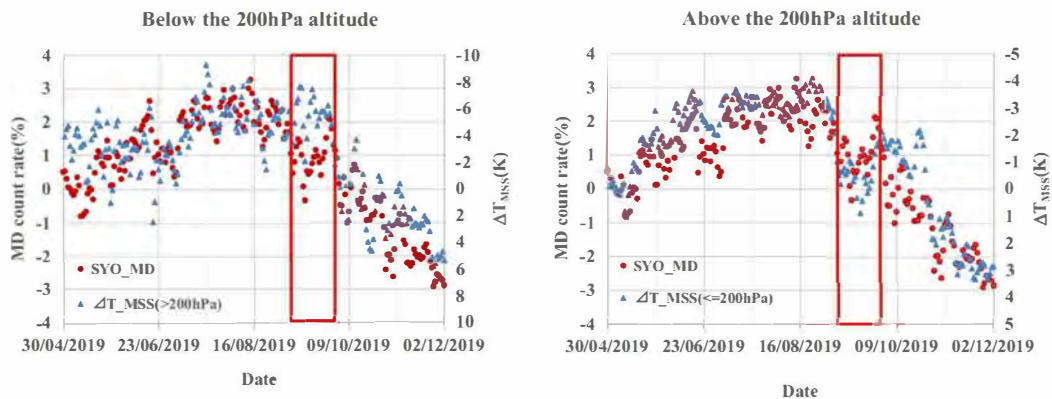

**Fig. 13.** Pressure corrected daily mean SYO_MD_V count rate shown by red solid circles and the left vertical axis. $\Delta T_{MSS}$ (calculated with Eq. (3)) is shown by blue solid circles on the right vertical axis using an inverted scale. The left panel is calculated from the temperature below the 200 hPa altitude and the right panel from the temperature above the 200 hPa altitude. The SSW event is indicated by a red rectangle in both panels. Minor tick marks on the horizontal axis are indicated every 27 days.

The pressure corrected hourly count rates recorded by Thule NM (THUL), South Pole NM (SOPO), and Syowa NM (SYO_NM) during 12 days between 20 August and 1 September, 2018 are shown in figure 14. The GCR intensity variation observed during this event by THUL at a northern high-latitude is quite different from the variations by SYO_NM and SOPO at southern highlatitudes. To separate the contributions from the GCR density and the north-south anisotropy, we also show averages and differences between THUL and SOPO and between THUL and SYO_NM. In the averages in panel (d), there is a decrease seen during the second half of 25 August corresponding to a Forbush decrease observed in the MFR, while in the difference in panel (e) a peak is seen during the first half of 26 August indicating the enhanced, north directed anisotropy. Also seen in panel (d) is a peak during the first half of 26 August indicating a GCR density exceeding the unmodulated level before the Forbush decrease. We also verified that the GCR variation observed by Syowa MD during this event is fairly consistent with the variation expected from the GCR density and anisotropy vector derived from the GMDN observation on hourly basis. We conclude that the Syowa NM and MD have successfully observed this event verifying our analyses of the GMDN data. This is a good example indicating that the observation at a single NM or MD is not enough for analyzing space weather events accurately and the network observation with multiple detectors plays an essential role.

# 5. Summary

New CR detectors (Syowa NM and Syowa MD) at Syowa Station started simultaneous and continuous observations of CR neutrons and muons, which are continuing. By analyzing the data collected in one and half years between February, 2018 and October, 2019, we showed that the seasonal variation of MD data due to the atmospheric temperature effect can be corrected by a method developed by Mendoça et al. (2016). Moreover, a decrease in CR muon count rate during the SSW event in September, 2019 is successfully recorded by Syowa MD. CR intensity variations observed by Syowa NM and MD during a space weather event in August, 2018 are consistent with other NM observations and the observation with the GMDN. We conclude that these new observations provide us with reliable and important data for space weather research and also for studying different responses of NM and MD to atmospheric/geomagnetic effects. Simultaneous observations with NM and MD at Syowa Station in Antarctic will play an essential role in the integrated analysis of data from the GMDN and the world-wide network of NMs.

A quick look website is online, which will help people in analyzing space weather events in near real time, and in studying the long term variations when the data are accumulated (see the URL in the footnote in Section 3).

Acknowledgements. The authors are grateful to Center for Antarctic Programs at NIPR (National Institute of Polar Research), JARE59 team, and Shirase crew for installing the system to Syowa Station. This research project is supported by NIPR, ISEE (Institute for Space-Earth Environmental Research), University of Delaware, and Shinshu University. Some of the scientific data of this research project has begun to be published at `http://polaris.nipr.ac.jp/~cosmicrays/` supported by ROIS-DS-JOINT2018. The Bartol Research Institute neutron monitor program is supported by the United States National Science Foundation under grants PLR-1245939 and PLR-1341562, and by Department of Physics and Astronomy and the Bartol Research Institute, the University of Delaware. The neutron monitor data from Thule are provided by the University of Delaware Department of Physics and Astronomy and the Bartol Research Institute. The neutron monitor data from South Pole and the South Pole Bares are provided by the University of Wisconsin, River Falls.

We acknowledge the NMDB database (`http://www.nmdb.eu`), founded under the European Union's FP7 program (contract no. 213007) for providing data.

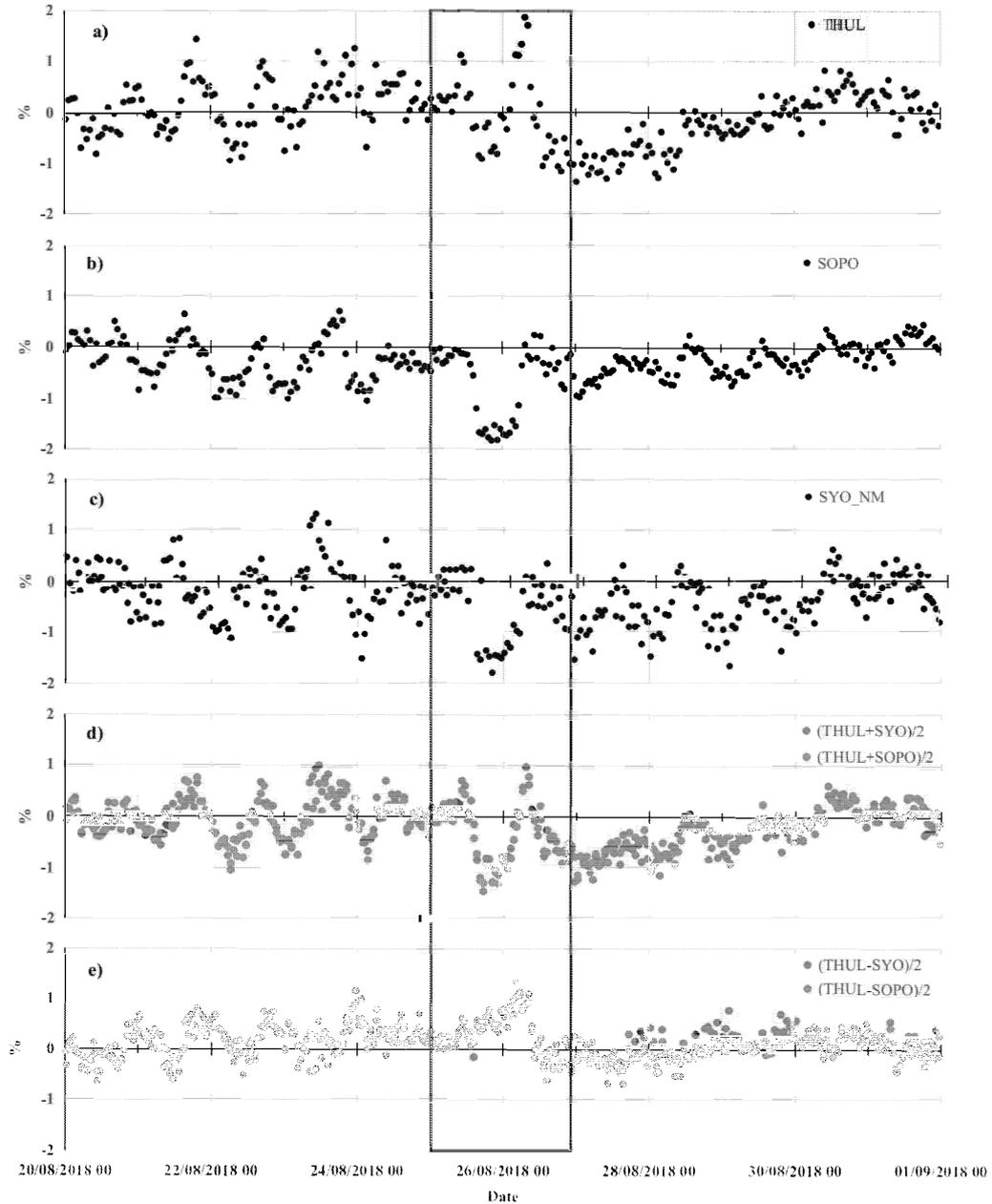

**Fig. 14.** Hourly count rate variations observed by a) Thule NM (THUL), b) South Pole NM (SOPO), and c) Syowa NM (SYO NM). NM at northern and southern high-latitudes show different variations. To see contributions from the GCR density and the north-south anisotropy separately, we plot the average and difference between NMs at northern and southern high-latitudes in panels d) and e). Orange and blue solid circles in panels d) and e) show the average and difference in THUL and SOPO pair and THUL and SYO NM pair, respectively. Tick marks on the horizontal axis are at one day intervals.